\documentclass[referee, pdflatex, sn-nature]{sn-jnl-modified}

\usepackage{graphicx}
\usepackage{multirow}
\usepackage{amsmath,amssymb,amsfonts}
\usepackage{amsthm}
\usepackage{mathrsfs}
\usepackage[title]{appendix}
\usepackage{xcolor}
\usepackage{textcomp}
\usepackage{manyfoot}
\usepackage{booktabs}
\usepackage{algorithm}
\usepackage{algorithmicx}
\usepackage{algpseudocode}
\usepackage{listings}
\usepackage{lineno}

\begin{document}

\title[Article Title]{Topological Surface Charge Detection via Terahertz Time-domain Spectroscopy}

\author[1]{\fnm{Tong} \sur{Shen}}\equalcont{These authors contributed equally to this work.}
\author[1]{\fnm{Yiyang} \sur{Xie}}\equalcont{These authors contributed equally to this work.}
\author[1]{\fnm{Yucheng} \sur{Dai}}\equalcont{These authors contributed equally to this work.}
\author[2]{\fnm{Yifan} \sur{Zhang}}
\author[1]{\fnm{Yuanze} \sur{Li}}
\author[2,3]{\fnm{Xufeng} \sur{Kou}}
\author*[1,4]{\fnm{Tian} \sur{Liang}}\email{tliang@mail.tsinghua.edu.cn}

\affil[1]{\orgname{State Key Laboratory of Low Dimensional Quantum Physics, Department of Physics, Tsinghua University}, \orgaddress{\city{Beijing} \postcode{100084}, \country{People's Republic of China}}}

\affil[2]{\orgname{School of Information Science and Technology, ShanghaiTech University}, \orgaddress{\city{Shanghai} \postcode{201210}, \country{People's Republic of China}}}

\affil[3]{\orgname{ShanghaiTech Laboratory for Topological Physics, School of Physical Science and Technology, ShanghaiTech University}, \orgaddress{\city{Shanghai} \postcode{201210}, \country{People's Republic of China}}}

\affil[4]{\orgname{Frontier Science Center for Quantum Information}, \orgaddress{\city{Beijing} \postcode{100084}, \country{People's Republic of China}}}

\abstract{

The topological magnetoelectric effect (TME) is a condensed-matter realization of the four-dimensional quantum Hall effect (4D-QHE), manifesting as quantized surface charge accumulation proportional to an applied magnetic field. To date, however, no optical technique has been developed to directly probe this charge accumulation. Here, we demonstrate a terahertz time-domain spectroscopy method for direct detection of surface charge accumulation—a physical quantity relevant to both the 4D-QHE and 2D-QHE, in sharp contrast to the previous optical measurements, which focused on the Hall conductivity $\sigma_{xy}$ of the 2D-QHE. Using a chromium-doped (Bi,Sb)$_2$Te$_3$ thin film, we achieve sub-milliradian Faraday rotation precision. Extending this scheme to axion insulators, we predict that the TME gives rise to an imaginary Faraday rotation linear in frequency, whose slope directly reflects the single-surface charge density. With further improvements in sample thickness and precision, this approach offers a viable pathway toward direct verification of the TME and 4D-QHE.

}

\maketitle

The topological magnetoelectric effect (TME) in three-dimensional topological insulators (3D TIs)~\cite{Kane2010,Qi2011} is a condensed-matter realization of the four-dimensional quantum Hall effect (4D QHE)~\cite{ZhangHu2001,QiHughesZhang2008,Essin2009}. It is described by the equation
\begin{equation}
\Delta P = \frac{e^2}{2h} N_{\mathrm{Ch}}^{(2)} \Delta B,
\end{equation}
where $N_{\mathrm{Ch}}^{(2)}$ is the second Chern number (an integer), and the four-dimensional parameter space is defined by the three spatial dimensions plus time. The TME occurs in a 3D TI when its surface magnetizations point uniformly either all inward or all outward, giving rise to an axion insulator (AI) state~\cite{Mogi2017,Mogi2017a,Xiao2018,Liu2020,Xu2014,Zhuo2023,Bai2024}, which is a single domain state of 4D QHE. In this state, a change in magnetic field $\Delta B$ produces a quantized change in parallel electric polarization $\Delta P$ throughout the bulk, determined by the factor $\frac{e^2}{2h} N_{\mathrm{Ch}}^{(2)}$. In a thin AI film under a perpendicular magnetic field, this polarization change characteristic of 4D QHE results in quantized surface charge accumulation of equal magnitude and opposite sign on the top and bottom surfaces ($\eta_\mathrm{4D-QHE} = \Delta B \cdot \frac{e^2}{2h} N_{\mathrm{Ch}}^{(2)}$ on one surface and $-\eta_\mathrm{4D-QHE}$ on the other). By contrast, in a multi-domain state of 4D QHE (where the bulk polarization change vanishes), i.e., the quantum anomalous Hall (QAH) state of a 3D TI, quantized surface charge accumulation of equal magnitude and same sign appears on the top and bottom surfaces, resulting in a net charge accumulation of $\eta_\mathrm{2D-QHE} = \Delta B \cdot \frac{e^2}{h}$, characteristic of 2D QHE. These arguments demonstrate that charge accumulation is relevant to both the 2D QHE and 4D QHE, whereas the Hall conductivity $\sigma_{xy}$ is strictly relevant to only the 2D QHE. 

Therefore, developing a technique which can directly access the charge accumulation is a crucial step toward detecting the 4D QHE. In transport, an out-of-plane capacitive coupling method for directly measuring the charge accumulation has already been developed~\cite{Li2025, Li2026}. To date, however, no optical technique has been developed to directly probe the charge accumulation. Here, we demonstrate a terahertz time-domain spectroscopy (THz-TDS) method for direct detection of charge accumulation. Operating in the THz regime is crucial, as it ensures that the photon energy remains below the surface band gap of $\sim 20-50$ meV~\cite{Tokura2016,Jiang2015,Lee2015}, preventing the interband excitations that would otherwise cause deviations from the ideal charge accumulation per unit magnetic field. This optical approach to detecting charge accumulation is applicable to a wider range of samples grown on thick substrates and with severe dissipation, in stark contrast to transport-based measurements as shown in Fig.~1(d) and Fig.~4(d). Importantly, even with $\alpha = 0.999$ active capacitive compensation—corresponding to nearly complete compensation—the transport-based signal remains more sensitive to finite longitudinal conductivity than the optical signal~\cite{Li2026}. Thus, while active capacitive compensation can significantly improve the performance of transport-based detection, our optical method retains a practical advantage because it directly probes the surface charge accumulation per unit magnetic field without requiring electrical contacts, precise capacitive compensation, or specific substrate thicknesses.

Quite generally, applicable both to the AI (4D QHE) and QAH (2D QHE) states, the THz-TDS yields a Faraday rotation angle that is directly proportional to the charge accumulation per unit magnetic field. We first consider the QAH state (more generally, a magnetically doped 3D TI where both the top and bottom surfaces yield the same sign of charge accumulation). As shown in the Supplemental Materials Section V, solving Maxwell's equations at the sample interfaces for the general case yields the charge accumulation per unit magnetic field
\begin{equation}
\frac{\eta}{B_{z}} = \left( \frac{\varepsilon_{2}}{n_{2}} + \frac{\varepsilon_{1}}{n_{1}}\frac{\cos\varphi_{2}}{\cos\varphi_{1}} \right) c \tan\theta_F,
\label{eq:eta_main}
\end{equation}
where subscripts~1 and~2 denote the substrate and vacuum, respectively (see Fig.~1(b)). Importantly, we made no assumption about the relation between the current and the electric field, so no information about the Hall conductivity is needed. In other words, Eq.~\ref{eq:eta_main} is applicable to general systems, including those where the Hall conductivity is not even defined. This clarifies the physical meaning of the Faraday rotation angle $\tan\theta_F$: it directly and independently detects the charge accumulation per unit magnetic field (up to a proportionality constant). This finding of directly detecting the charge accumulation via Faraday rotation is new and has not been obtained previously. We note that the technique of THz-TDS and Faraday rotation measurement itself is widespread and has recently been theoretically proposed~\cite{Tse2010PRL,Tse2010PRB,Tse2011PRB} and experimentally applied to electro-optic studies of topological insulators~\cite{Wu2016,Tokura2016,Shimano2010,Shimano2013, neu2018tutorial}. However, the physical quantity theoretically proposed or experimentally extracted in those previous studies is the Hall conductivity, a physical quantity accessible only to 2D QHE, in sharp contrast to the charge accumulation, a physical quantity relevant to both the 2D QHE and 4D QHE.
We demonstrate this optical method of detecting the surface charge accumulation on a chromium-doped $\mathrm{(Bi,Sb)_2Te_3}$ thin film grown by molecular beam epitaxy (MBE) on a GaAs substrate~\cite{Ji2022}, which is transparent to THz light. 
The surface charge accumulation is quantified through high-precision Faraday rotation measurements with sub-milliradian resolution.

The chromium-doped $\mathrm{(Bi,Sb)_2Te_3}$ thin film~\cite{Chang2013,Ji2022} (a 2D QHE system) shares the same surface state physics as the axion insulator state (a 4D QHE system), and therefore our optical detection scheme bridges the 2D QHE and 4D QHE, as described below.

Fig.~1(a) illustrates the THz-TDS setup~\cite{neu2018tutorial} used for Faraday rotation measurements. A femtosecond laser drives photoconductive antennas for THz generation and coherent detection. A mechanical delay line maps the time-domain electric field waveform, which is read out via a lock-in amplifier. Wire-grid polarizers P1 and P3 guarantee linear polarization of the emitted and detected beams, while the rotatable polarizer P2 selects the transmitted components parallel ($E_X$) and perpendicular ($E_Y$) to the incident polarization. 
The sample is mounted in an OptiCool cryostat with optical windows for independent control of temperature and magnetic field. Further experimental details are described in the Supplemental Material.

The sample is fixed on a tilted holder so that the angle between its normal and the incident beam is $\varphi_0$. 
As sketched in Fig.~1(b), two coordinate systems describe the measurement geometry. 
In the laboratory frame $(X,Y,Z)$, the THz pulse propagates along $Z$, the incident electric field $E_0$ is polarized along $X$, and after traversing the substrate/sample structure the field acquires a Faraday rotation defined by $\tan\theta_F = E_Y/E_X$. 
The local frame $(x,y,z)$ is attached to the sample with $z$ along its normal and $x$ coinciding with the incident polarization ($X$). 
In this frame, the obliquely incident pulse possesses a magnetic field component $B_{z}$ along $z$, which gives rise to a non-trivial surface charge accumulation $\eta$. 
Before each measurement, a DC magnetic field $B_{\mathrm{DC}}=\pm1\,\mathrm{T}$ is applied to magnetize the sample and then reduced to zero for data acquisition; antisymmetrizing the signals obtained under opposite polarities of $B_{\mathrm{DC}}$ eliminates any Faraday rotation not originating from the topological surface state.

A representative time-domain trace recorded at $2\,\mathrm{K}$ with $\varphi_0=45^{\circ}$ is shown in Fig.~1(c). 
$E_X$ is typically three orders of magnitude larger than $E_Y$, and our setup achieves a precision better than $1\,\mathrm{mrad}$ in the determination of the rotation angle.

\begin{figure*}[htbp]
  \centering
  \includegraphics[width=1\linewidth]{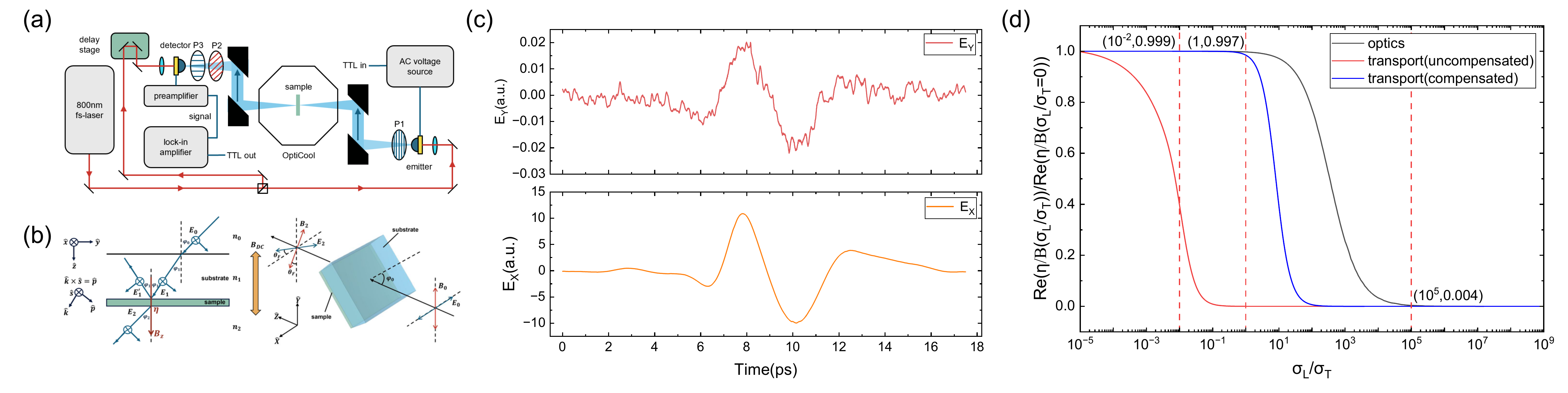}
  \caption{\textbf{THz time-domain spectroscopy and charge accumulation.}
  (a) Experimental setup for Faraday rotation measurements. The THz-TDS system is integrated with a lock-in amplifier, synchronized via TTL signals with the AC voltage source. Polarizers P1 and P3 fix the polarization of the emitter and detector, respectively, while the rotatable polarizer P2 selects the transmitted components $E_X$ or $E_Y$. The sample is mounted in an OptiCool cryostat with optical windows for temperature and magnetic field control.
  (b) Schematic of the Faraday rotation geometry in a magnetic topological insulator. A linearly polarized THz pulse propagates through the substrate/sample structure, producing transmitted components and a finite rotation angle $\theta_F$.
  (c) Time-domain waveforms measured at $2\,\mathrm{K}$ with an incident angle of $45^{\circ}$. $E_X$ and $E_Y$ are the components parallel and perpendicular to the incident polarization.
  (d) Normalized field-induced charge accumulation as a function of $\sigma_L/\sigma_T$ for the optical (black), uncompensated transport (red), and actively compensated transport (blue) detection schemes. The optical signal is robust against finite $\sigma_L$, decaying by less than $0.3\%$ even when $\sigma_L \sim \sigma_T$, whereas the transport signal vanishes rapidly. Active capacitive compensation substantially improves the transport response, but is still more sensitive to $\sigma_L$ than optical method; the compensated curve is calculated with $\alpha=0.999$. The transport curves are simulated for an AC frequency of $1\,\mathrm{kHz}$, a top-gate thickness of $200\,\mathrm{nm}$ with $\varepsilon_r=9$, and a sample radius of $1000\,\mu\mathrm{m}$.}
  \label{fig:fig1}
\end{figure*}

\begin{figure*}[htbp]
  \centering
  \includegraphics[width=1\linewidth]{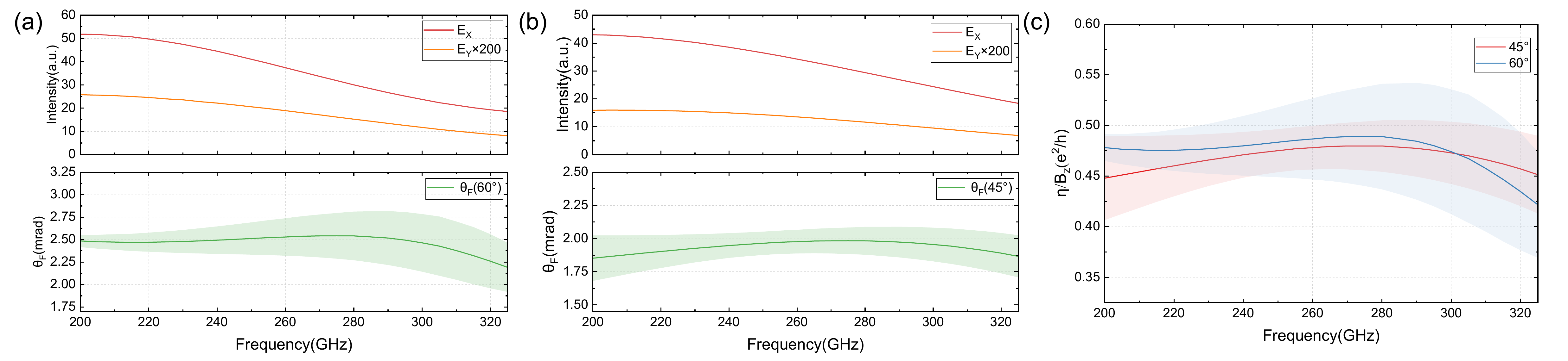}
  \caption{\textbf{Angular dependence of Faraday rotation and charge accumulation.} 
  Intensity spectra and corresponding Faraday rotation $\theta_F$ measured at $2\,\mathrm{K}$ for incident angles of $60^{\circ}$ (a) and $45^{\circ}$ (b). Shaded regions indicate the uncertainty in $\theta_F$.
  (c) Surface charge accumulation $\eta/B_z$ extracted from $\theta_F$ for $60^{\circ}$ (blue) and $45^{\circ}$ (red), using $n_1=3.5$ for the GaAs substrate and $n_2=1$ for air. The shaded areas represent the propagated uncertainty obtained from the standard error of the mean (SEM) of repeated measurements.}
  \label{fig:fig2}
\end{figure*}

After Fourier transforming the time-domain signals $E_X(t)$ and $E_Y(t)$, we obtain the complex spectra $E_X(\omega)$ and $E_Y(\omega)$. 
The Faraday rotation spectrum is then given by
\begin{equation}
\tan\theta_F(\omega) = \operatorname{Re}\!\left[\frac{E_Y(\omega)}{E_X(\omega)}\right].
\end{equation}

Figs.~2(a) and~2(b) show the intensity spectra and corresponding $\theta_F(\omega)$, and Fig.~2(c) shows the charge accumulation per unit magnetic field $\eta/B_z$ obtained from Eq.~\eqref{eq:eta_main}  at $2\,\mathrm{K}$ for incident angles of $60^{\circ}$ and $45^{\circ}$, respectively, with uncertainties indicated by shading. 

The experimentally obtained ratio of the charge accumulation per unit magnetic field at the two incident angles at $250\,\mathrm{GHz}$  is $\frac{\eta}{B_{z}}(60^{\circ})/\frac{\eta}{B_{z}}(45^{\circ}) = 1.018 \approx 1$, showing essentially no angular dependence. We note that so far we have not made any assumptions about the relation between the current $\vec j$ and the electric field $\vec E$, and in particular, no information about the Hall conductivity is needed to obtain the above result.

Now by using the relation $\vec j = \boldsymbol\sigma \vec E$ with $\boldsymbol\sigma = \begin{pmatrix}
 \sigma_L & \sigma_T\\
 -\sigma_T & \sigma_L
\end{pmatrix}$ applicable to the chromium-doped $\mathrm{(Bi,Sb)_2Te_3}$ thin film, Eq.~\eqref{eq:eta_main} reduces to (see SM Sec.~VI for details)
\begin{equation}
\frac{\eta}{B_{z}} = \sigma_{T}\left( 1 - \frac{\sigma_{L}}{\sigma_{T}} \cos\varphi_{2} \tan\theta_F \right).
\label{eq:eta_sigma}
\end{equation}
Because $\theta_F$ is on the order of $1\,\mathrm{mrad}$, the second term is negligible even for finite $\sigma_L \sim \sigma_T$, explaining the experimentally observed angular dependence of the charge accumulation per unit magnetic field. This result also confirms that a finite longitudinal conductivity does not compromise the optical measurement.

\begin{figure}[htbp]
  \centering
  \includegraphics[width=1\linewidth]{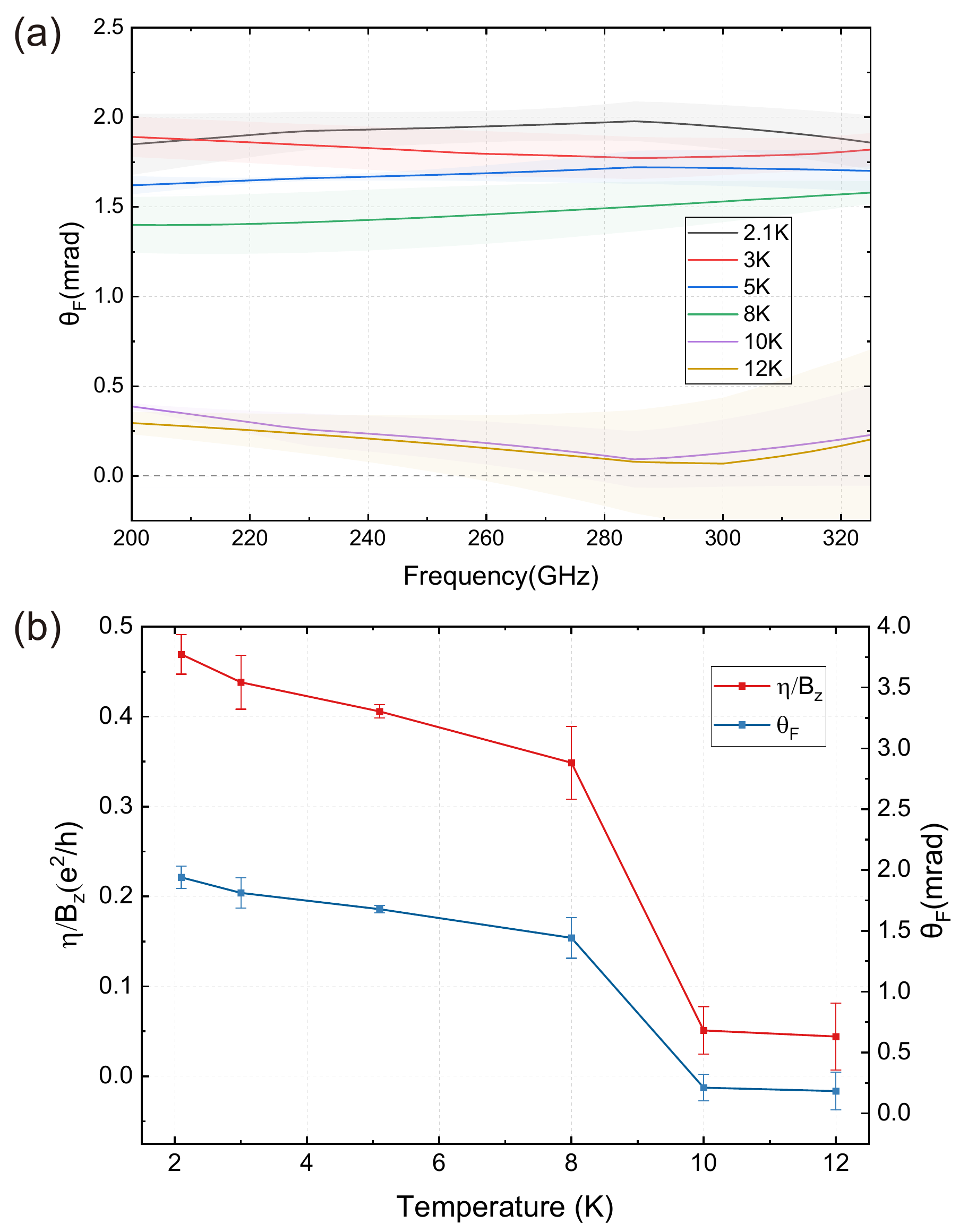}
  \caption{\textbf{Temperature dependence of Faraday rotation and charge accumulation.} 
  (a) Faraday rotation spectra $\theta_F(\omega)$ measured at an incident angle of $45^{\circ}$ for several temperatures. Shaded areas indicate the uncertainty.
  (b) Temperature dependence of $\theta_F$ (right axis) and the extracted charge accumulation $\eta/B_{z}$ (left axis) at $250\,\mathrm{GHz}$.}
  \label{fig:fig3}
\end{figure}

We further investigated the temperature response of the charge accumulation. 
At each temperature, the sample was magnetized following the same antisymmetrization procedure described above. 
Fig.~3(a) displays $\theta_F(\omega)$ recorded at $\varphi_0 = 45^{\circ}$ for a set of temperatures. 
The rotation angle decreases monotonically with increasing temperature and becomes essentially undetectable above $10\,\mathrm{K}$. 
We obtain $\eta/B_{z}$ at $250\,\mathrm{GHz}$ using Eq.~\eqref{eq:eta_main}, as shown in Fig.~3(b). 
The extracted charge accumulation decays as temperature increases and $\sigma_T$ decreases. The fact that charge accumulation can be detected at elevated temperatures, where $\sigma_L/\sigma_T$ becomes large (see Fig.~S4), further underscores that the optical method is immune to the influence of $\sigma_L$.

\begin{figure*}[htbp]
  \centering
  \includegraphics[width=1\linewidth]{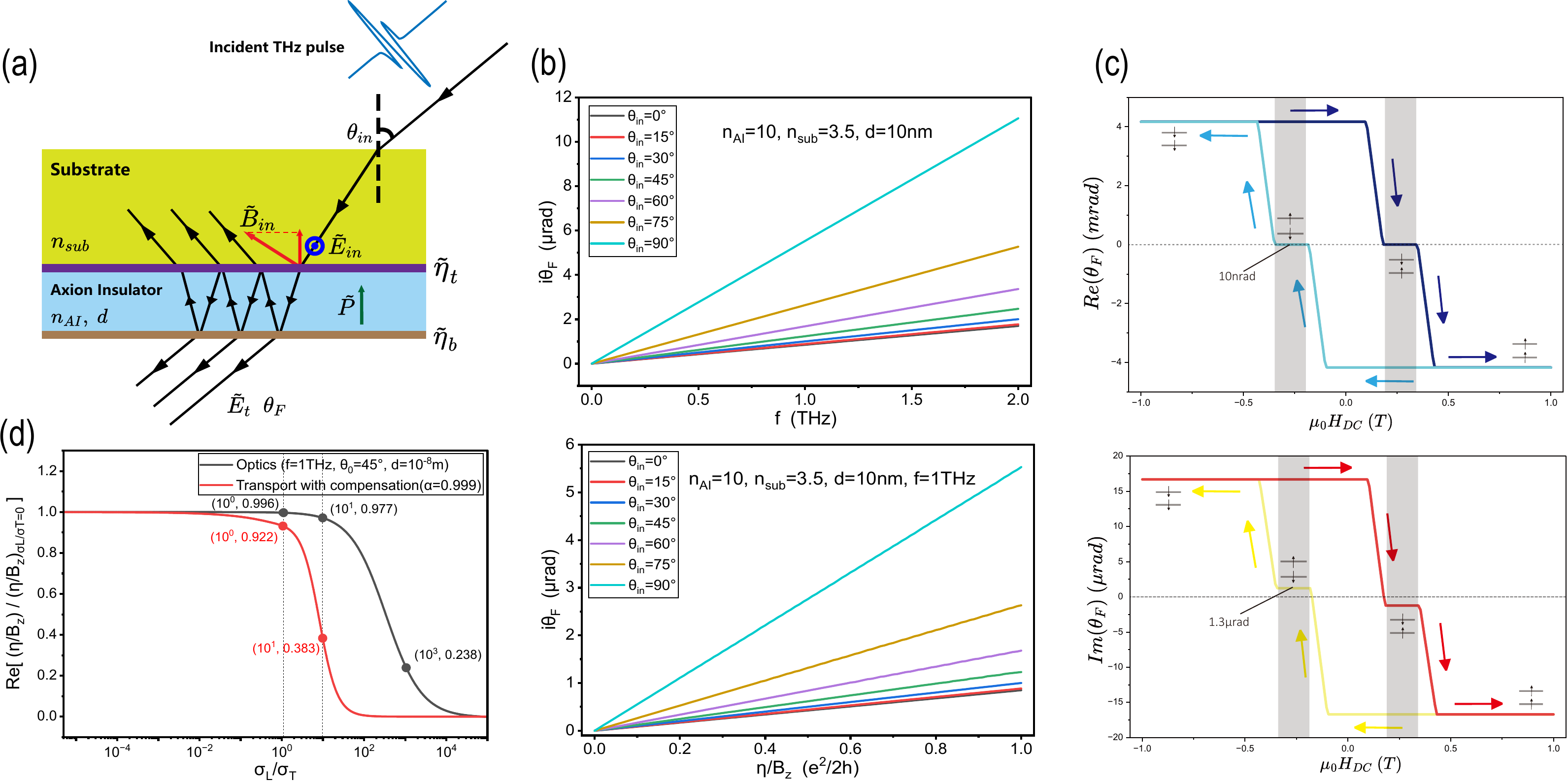}
  \caption{\textbf{Optical detection of the topological magnetoelectric effect.}
  (a) Schematic of probing the TME via Faraday rotation. A THz pulse impinges from vacuum at an angle $\theta_0$, passes through a substrate of refractive index $n_{\mathrm{sub}}$, and undergoes multiple-beam interference inside an axion insulator of thickness $d$ and refractive index $n_{\mathrm{AI}}$.
  (b) Frequency dependence of the imaginary part of the Faraday rotation angle of transmitted light at different simulated incident angles, and the relationship between the imaginary part of the Faraday rotation angle and the single surface charge accumulation per unit magnetic field at different simulated incident angles. (c) Simulated dependence of the real and imaginary parts of the Faraday rotation angle on the DC magnetic field when $ \theta_0=45^{\circ}$, with the same parameter settings as in (b). The shaded region indicates the axion insulator regime. (d) Normalized charge accumulation  per unit magnetic field of the axion insulator as a function of $\sigma_L/\sigma_T $ (optical: black; transport: red). In TME measurements, the optical method remains robust against relatively high longitudinal conductance. }
  \label{fig:fig4}
\end{figure*}

Having successfully detected the charge accumulation of the topological surface state, we now apply the same optical approach to simulate the topological magnetoelectric effect (TME) in an axion insulator~\cite{Mogi2017,Mogi2017a,Xiao2018,Liu2020,Xu2014,Zhuo2023,Bai2024}. 
Fig.~4(a) sketches the measurement configuration: an $s$-polarized THz wave impinges at a finite angle, and its perpendicular magnetic field component induces an out-of‑plane electric polarization difference $\Delta P = \frac{e^2}{2h} N_{\mathrm{Ch}}^{(2)} \Delta B$, which couples back to the beam and produces Faraday and Kerr rotations.

The multiple reflections inside the thin axion insulator provide a direct route to model this coupling. 
Because the sample thickness $d$ is much smaller than the THz wavelength, it behaves as a Fabry‑Pérot cavity; the beam undergoes repeated partial reflections at the two surfaces that carry opposite TME‑induced charge densities, and the outgoing waves interfere coherently. 
Each interface can be described by a non‑diagonal reflection–transmission matrix. 
Although this matrix method does not yield compact analytical formulas, it provides exact numerical simulations.

A simpler physical picture emerges from the phase difference between the two surfaces. 
Owing to the finite thickness $d$ and refractive index $n_{\mathrm{AI}}$, the magnetic fields at the top and bottom surfaces acquire a small phase lag $\phi$. 
Consequently, the originally opposite surface charges $\eta_t = \eta$ and $\eta_b = -\eta$ become $\eta_t = \eta$ and $\eta_b = -\eta e^{i\phi}$. 
The net charge accumulation in the axion insulator is then
\begin{equation}
\eta_{\mathrm{total}} = \eta_t + \eta_b = \eta(1-e^{i\phi}) = \eta(1-\cos\phi - i\sin\phi),
\end{equation}
where the imaginary part signifies a $\pi/2$ phase shift with respect to the driving magnetic field. 
For $\phi \ll 1$, the real part is negligible compared to the imaginary part, giving
\begin{equation}
\eta_{\mathrm{total}} \approx -i\phi\,\eta = -i\frac{2\pi n_{\mathrm{AI}} f d }{c\cos\theta_{\mathrm{AI}}}\eta,
\label{eq:eta_total}
\end{equation}
with $\theta_{\mathrm{AI}}$ the propagation angle inside the axion insulator and $f$ the frequency. 
Analogously to the QAH case, this imaginary net charge rotates the polarization plane as
\begin{equation}
\frac{\eta_{\mathrm{total}}}{B_z} = \varepsilon_0\, \kappa(n_{\mathrm{sub}}, n_{\mathrm{AI}}, \theta_0)\,c\,\tan\theta_F,
\label{eq:TME_rotation}
\end{equation}
where $\kappa$ is a complex coefficient encapsulating the multiple-beam interference and depends on the refractive indices and incident angle; its exact value is obtained numerically. 
Equations~\eqref{eq:eta_total} and \eqref{eq:TME_rotation} reveal that the Faraday rotation is proportional to the single-surface TME charge density per unit magnetic field and to the frequency. 
The rigorous numerical simulations in Fig.~4(b) confirm this linear relationship between the imaginary part of $\theta_F$, the frequency $f$, and the single-surface charge density per unit magnetic field $\eta/B_z$.

To experimentally verify the TME, one measures the transmitted THz waveform, performs Fourier analysis, and extracts the frequency‑dependent complex Faraday rotation $\theta_F(\omega) = \theta_F'(\omega) + i\theta_F''(\omega)$. As shown in Fig.~4(c), two signatures must be observed: (i) the real part $\theta_F'$ is essentially zero (below $10\,\mathrm{nrad}$), indicating that the charges on the two surfaces are opposite and thereby produce no in‑phase net charge with respect to the magnetic field. Therefore, if a finite real-part Faraday rotation angle is observed, the symmetric charge can be calculated accordingly, thereby excluding interference from charge accumulation unrelated to the TME; (ii) the imaginary part $\theta_F''$ exhibits a nonzero slope with frequency. 
This slope is directly proportional to the single‑surface charge density per unit magnetic field, so measuring it yields the TME‑induced charge density per unit magnetic field on one surface of the axion insulator. Importantly, even if the sample is incomplete and has asymmetry in the top and bottom surface charges, by monitoring the real-part signals in addition to the imaginary part signals of the Faraday measurements, one can obtain the correct single‑surface charge density per unit magnetic field originating from the TME (see supplemental materials for more information).

The optical method's advantage over transport-based methods for detecting the TME lies in its strong robustness against finite longitudinal conductance $\sigma_L$ in the sample and its wide range of applicability with respect to the substrate thickness on which the sample is grown. Fig.~4(d) shows that when $\sigma_L \sim \sigma_T$, the TME signal in the optical method sustains a signal attenuation of less than 1\%, implying that samples with relatively large longitudinal conductance can still produce the TME signal. At the same time, increasing the sample thickness significantly enhances the strength of the rotation signal. Since the Faraday rotation angle is approximately linear with thickness, if the sample thickness reaches $100\,\mathrm{nm}$~\cite{Zhuo2023}, the theoretical rotation angle will be on the order of $10\,\mu\mathrm{rad}$, making it possible for this rotation signal to be observed by high‑precision measurement devices. 

In conclusion, we have for the first time recognized the true physical meaning of Faraday rotation---charge accumulation per unit magnetic field (up to a proportionality constant), applicable to both 4D QHE and 2D QHE systems in a unified way, and validated it through angular-dependent measurements of the surface charge accumulation of a chromium-doped $\mathrm{(Bi,Sb)_2Te_3}$ thin film (magnetic topological insulator) using THz-TDS. 
Under oblique incidence, the Faraday rotation angle agrees with theoretical predictions, and the measurement precision reaches sub-milliradian.
A key advantage of this optical method is that it remains applicable even for materials with sizable longitudinal conductivity $\sigma_L$ and thick substrates.

Since the charge-accumulation mechanism in the QAH (2D QHE) regime closely parallels that of the TME (4D QHE), our work constitutes an important step toward the direct optical detection of the TME. 
We further show that the TME manifests as a frequency-linear Faraday rotation, whose slope directly reflects the charge accumulation per unit magnetic field on a single surface. 
With feasible improvements in sample thickness and measurement precision shown in the supplemental materials, we anticipate that the optical measurement of the TME will become experimentally realizable.

T.L. acknowledges the support for the project by the National Key R$\&$D Program of China (No. 2021YFA1401600). We thank Chong Wang and Luyi Yang's group for helpful discussions.

T.L. conceived and designed the project, provided overall direction, supervised the experiments, and coordinated collaborations among the research groups. T.S., Y.X., and Y.D. built from scratch the oblique incident terahertz time-domain spectroscopy system. T.S. and Y.X. performed the optical measurements and analyzed the experimental data. T.S., Y.X., and Y.D. carried out theoretical analyses and numerical simulations with in-depth discussions with T.L. Y.Z. grew the magnetic topological-insulator thin films under the supervision of X.K. Y.L. performed the transport measurements and related numerical simulations. T.S., Y.X., and T.L. analyzed the results and wrote the manuscript with input from all authors. All authors discussed the results and provided feedback on the manuscript.

\bibliography{MainPaper}

\end{document}